\documentclass[lettersize,journal]{IEEEtran}
\usepackage{amsmath,amsfonts}
\usepackage{algorithmic}
\usepackage{algorithm}
\usepackage{array}
\usepackage[caption=false,font=normalsize,labelfont=sf,textfont=sf]{subfig}
\usepackage{textcomp}
\usepackage{stfloats}
\usepackage{url}
\usepackage{verbatim}
\usepackage{graphicx}
\usepackage{cite}
\begin{document}

\title{D\textsuperscript{2}IP: Deep Dynamic Image Prior for 3D Time-sequence Pulmonary Impedance Imaging}

\author{Hao Fang, \IEEEmembership{Graduate Student Member, IEEE}, Hao Yu, Sihao Teng, \IEEEmembership{Graduate Student Member, IEEE}, Tao Zhang, Siyi Yuan, Huaiwu He, Zhe Liu, \IEEEmembership{Member, IEEE}, Yunjie Yang, \IEEEmembership{Senior Member, IEEE} 
\thanks{Hao Fang, Hao Yu, Sihao Teng, Zhe Liu and Yunjie Yang are with the SMART Group, Institute for Imaging, Data and Communications, School of Engineering, The University of Edinburgh, Edinburgh, UK. (Correspondence authors: Yunjie Yang and Zhe Liu; Email: y.yang@ed.ac.uk and zz.liu@ed.ac.uk).}
\thanks{Tao Zhang is with the Department of Intensive Care Unit, Tianjin Huanhu Hospital, Tianjin, China.}
\thanks{Siyi Yuan and Huaiwu He are with the State Key Laboratory of Complex Severe and Rare Diseases, Department of Critical Care Medicine, Peking Union Medical College, Peking Union Medical College Hospital, Chinese Academy of Medical Sciences, Beijing, China.}
\thanks{This work was supported in part by Noncommunicable Chronic Diseases-National Science and Technology Major Project (2023ZD0512000) and Prunus Medical-Edinburgh joint research fund.}}

\markboth{Journal of \LaTeX\ Class Files,~Vol.~14, No.~8, August~2021}%
{Shell \MakeLowercase{\textit{et al.}}: A Sample Article Using IEEEtran.cls for IEEE Journals}

\IEEEpubid{0000--0000/00\$00.00~\copyright~2021 IEEE}

\maketitle

\begin{abstract}
Unsupervised learning methods, such as Deep Image Prior (DIP), have shown great potential in tomographic imaging due to their training-data-free nature and high generalization capability. However, their reliance on numerous network parameter iterations results in high computational costs, limiting their practical application, particularly in complex 3D or time-sequence tomographic imaging tasks. To overcome these challenges,  we propose Deep Dynamic Image Prior ($D^2IP$), a novel framework for 3D time-sequence imaging. $D^2IP$ introduces three key strategies—Unsupervised Parameter Warm-Start (UPWS), Temporal Parameter Propagation (TPP), and a customized lightweight reconstruction backbone, 3D-FastResUNet—to accelerate convergence, enforce temporal coherence, and improve computational efficiency. Experimental results on both simulated and clinical pulmonary datasets demonstrate that $D^2IP$ enables fast and accurate 3D time-sequence Electrical Impedance Tomography (tsEIT) reconstruction. Compared to state-of-the-art baselines, $D^2IP$ delivers superior image quality—with a 24.8\% increase in average MSSIM and an 8.1\% reduction in ERR—alongside significantly reduced computational time (7.1× faster), highlighting its promise for clinical dynamic pulmonary imaging.
\end{abstract}


\section{Introduction}
\label{sec:introduction}
\IEEEPARstart{P}{ulmonary} imaging plays a critical role in the early diagnosis, monitoring, and management of respiratory diseases such as pulmonary edema, chronic obstructive pulmonary disease (COPD), and acute respiratory distress syndrome (ARDS) \cite{b47,b48,b49,b50}. Among available techniques, tomographic imaging modalities—including Computed Tomography (CT) \cite{b51,b52} and Magnetic Resonance Imaging (MRI) \cite{b53,b54}—are widely used in clinical pulmonary imaging due to their ability to produce high-resolution anatomical images. These methods reconstruct internal anatomical structures by solving ill-posed inverse problems that map indirect measurements to tissue properties.

Despite their diagnostic value, CT and MRI are not suitable for continuous or bedside pulmonary monitoring. CT involves ionizing radiation, while both CT and MRI suffer from low temporal resolution, high operational costs, and limited portability. These limitations hinder their use in bedside, real-time imaging and assessments. Electrical Impedance Tomography (EIT) has emerged as a promising alternative for dynamic pulmonary imaging \cite{b22,b23,b24,b46}. EIT is radiation-free, low-cost, portable, and capable of high temporal resolution, offering a clinically viable solution for continuous bedside monitoring. It reconstructs conductivity distributions by injecting small currents and measuring resulting boundary voltages \cite{b20,b21}. However, the inherently ill-posed nature of EIT severely limits its spatial resolution, especially in noisy and complex clinical environments. 

To mitigate this, prior information is commonly incorporated to regularize the inverse problem. Classical handcrafted priors—such as Tikhonov regularization \cite{b9}, $\ell_1$-norm \cite{b10}, and Total Variation (TV) \cite{b11}—can stabilize ill-posed inverse problems but often require careful parameter tuning and degrade under strong noise or highly ill-posed conditions. Supervised deep learning (DL) methods have shown promise in learning complex inverse mappings from data, particularly for single-frame EIT reconstruction. Representative methods—such as convolutional neural networks (CNNs) \cite{b30,b31,b32}, variational autoencoders (VAEs) \cite{b33}, and generative adversarial networks (GANs) \cite{b34}—have achieved promising results by leveraging large-scale labeled datasets and data-driven priors. However, their reliance on expensive, task-specific training data significantly limits generalizability, especially in complex clinical settings involving unseen anatomical structures or variable imaging conditions \cite{b15}.

In response, unsupervised learning methods have recently gained traction in EIT due to their training-data-free nature and strong generalization capability. D Liu et al. \cite{b35} introduced Deep Image Prior (DIP) into EIT by using a U-Net as an implicit regularizer, enabling stable and high-quality 2D reconstruction by reformulating the task as a neural network parameter optimization problem. Z. Liu et al. \cite{b36} proposed Regularized Shallow Image Prior (R-SIP), which incorporates handcrafted priors into a shallow three-layer multilayer perceptron (MLP) for robust 2D and 3D EIT reconstructions. 

However, these methods primarily focus on single-frame EIT reconstruction, ignoring temporal correlations across frames. Multi-frame EIT reconstruction methods, including multi-frequency EIT (mfEIT) \cite{b37} and time-sequence EIT (tsEIT) \cite{b38}, have been developed to incorporate additional measurement dimensions, improving image quality and robustness. Among them, mfEIT exploits inter-frequency correlations to improve conductivity contrast and material differentiation. Recently, DL-based mfEIT approaches have been prevailing. For instance, the model-based learning framework MMV-Net \cite{b39} and the end-to-end learning algorithm SFCF-Net \cite{b40} can achieve high-quality 2D multi-frequency EIT reconstruction on circular phantoms. However, the data-driven nature of these methods limits their generalization ability, making them ineffective for reconstructing unseen samples and imaging regions \cite{b15}. MAIP \cite{b15} introduced an unsupervised framework for mfEIT to address this, but similar advancements for tsEIT remain limited. tsEIT has particular relevance for dynamic monitoring, yet constructing large-scale 3D tsEIT datasets is prohibitively resource-intensive due to clinical, anatomical, and ethical constraints. 
Unsupervised DL-based methods present a compelling solution for tsEIT by eliminating the need for labeled training data. However, methods like DIP typically require a large number of network iterations to achieve high-quality reconstructions, which results in high computational costs and limits their practicality for 3D time-sequence imaging tasks.

To address these challenges, we propose Deep Dynamic Image Prior ($D^2IP$), a novel unsupervised framework for efficient 3D time-sequence image reconstruction. $D^2IP$ integrates three key components to improve convergence, coherence, and efficiency: Unsupervised Parameter Warm-Start (UPWS), Temporal Parameter Propagation (TPP), and \emph{3D-FastResUNet}. The UPWS strategy initializes network parameters using the optimized parameters from a previously solved reconstruction task, significantly reducing the number of iterations. The TPP strategy enforces temporal continuity by propagating model weights across sequential frames. The \emph{3D-FastResUNet} is specifically designed to incorporate depthwise separable convolutions to further reduce the computational cost during the inference process. Together, these strategies allow $D^2IP$ to achieve a synergistic balance between computational efficiency, temporal consistency, and reconstruction accuracy for 3D time-sequence tomographic imaging. 

The key contributions of this paper are as follows:
\begin{itemize}
    \item We propose Deep Dynamic Image Prior ($D^2IP$), the first unsupervised DL-based framework for 3D time-sequence tomographic imaging. $D^2IP$ leverages implicit regularization without requiring labeled training data, significantly improving generalization and adaptability across diverse imaging scenarios.
    
    \item We introduce two key iteration-acceleration strategies—Unsupervised Parameter Warm-Start (UPWS) and Temporal Parameter Propagation (TPP) to enable fast convergence and ensure temporal coherence across sequential frames. 
    
    \item We design \emph{3D-FastResUNet}, a lightweight network architecture with depthwise separable convolutions to reduce inference cost. We also incorporate 4D Total Variation (4D-TV) regularization to further enhance spatio-temporal smoothness and reconstruction stability.
\end{itemize}

\section{3D Time-sequence EIT reconstruction}
The objective of 3D tsEIT is to reconstruct a sequence of 3D conductivity distributions that capture both spatial and temporal variations, based on sequential voltage measurements. The tsEIT forward model can be approximately linearized and expressed as:
\begin{equation}
\mathbf{V} = \mathbf{J\pmb{\Sigma}},
\label{e1}
\end{equation}
where $\mathbf{V}=[\Delta\mathbf{v}_{1},\Delta\mathbf{v}_{2},\ldots,\Delta\mathbf{v}_{i},\ldots,\Delta\mathbf{v}_{T}]\in\mathbb{R}^{M\times T}$, and $\mathbf{\Sigma}=[\Delta\pmb{\sigma}_{1},\Delta\pmb{\sigma}_{2},\ldots,\Delta\pmb{\sigma}_{i},\ldots,\Delta\pmb{\sigma}_{T}]\in\mathbb{R}^{Q\times T}$ represent the time-sequence of normalized voltage measurements and conductivity variations, respectively. $\mathbf{J}\in\mathbb{R}^{M\times V}$ represents the sensitivity tensor, which has been both normalized \cite{b41} and projected \cite{b15}. $i=1,2,...,T$ denotes the $i\text{-th}$ observation frame, and $T$ stands for the total number of time-sequence frames. $Q =(R\times C\times P)$ is the total number of voxels. $M$ denotes the number of voltage measurements per frame, while $R$, $C$, and $P$ represent the spatial dimensions (rows, columns, and planes) of each 3D reconstructed EIT image.

Due to the ill-posed nature of EIT, directly solving \eqref{e1} is highly underdetermined and sensitive to noise. To address this, we formulate the reconstruction as a regularized optimization problem. For each frame $i$, the conductivity change $\Hat{\Delta\pmb{\sigma}_i}$ is reconstructed by solving:
\begin{align}
\Hat{\Delta\pmb{\sigma}_i} = \arg\min_{\Delta\pmb{\sigma}_i} & \quad \|\Delta\mathbf{v}_i - \mathbf{J} \Delta\pmb{\sigma}_i\|  \notag + \lambda \mathcal{R}(\{\Hat{\Delta\pmb{\sigma}_j}\}_{j=1}^{i-1},\Delta\pmb{\sigma}_i), \\
&\quad \forall i \in \{1, 2,\dots,T\}.
\label{eq:argmin}
\end{align}
where $\|\cdot\|$ denotes a selected norm for data fidelity, such as the $\ell_1$, $\ell_2$, or \emph{Smooth}-$\ell_1$. $\mathcal{R}:~\mathbb{R}^{(R\times C\times P)\times i} \rightarrow \mathbb{R}$ represents the handcrafted regularization function.

\section{Methodology}
This section introduces the proposed Deep Dynamic Image Prior ($D^2IP$) framework for 3D time-sequence tomographic imaging. We first present the overall architecture of $D^2IP$ and then describe its three key strategies, i.e., Unsupervised Parameter Warm-Start (UPWS), Temporal Parameter Propagation (TPP), and \emph{3D-FastResUNet}. Finally, we introduce 4D Total Variation (4D-TV), which acts as a spatio-temporal regularizer to further improve reconstruction quality.

\subsection{Deep Dynamic Image Prior}
\begin{figure}[!t]
\centerline{\includegraphics[scale=0.44]{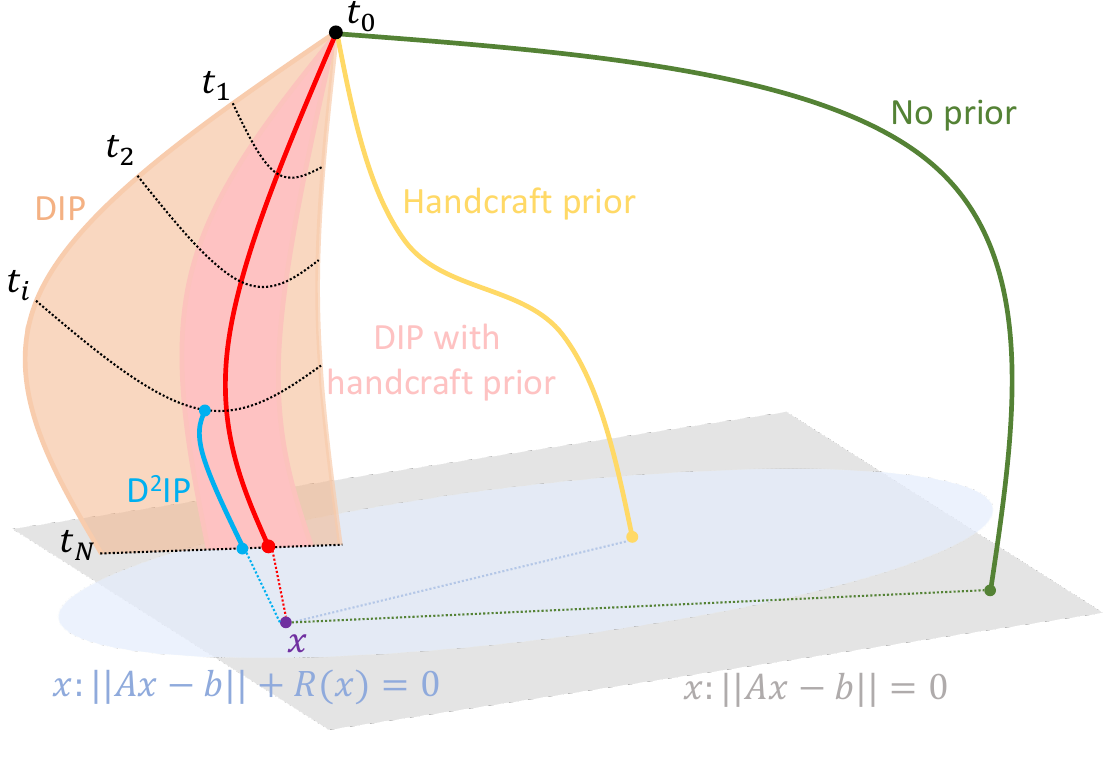}}
\caption{
Geometric interpretation of different reconstruction strategies in the solution space. The gray plane denotes the solution manifold of pure data fidelity, i.e., $\{x ~:~ \|Ax - b\| = 0\}$, while the light blue surface represents the regularized solution space with handcrafted prior, i.e., $\{x ~:~ \|Ax - b\| + R(x) = 0\}$. The green trajectory shows unconstrained optimization without prior, which deviates from the ground truth (purple dot). The yellow path corresponds to handcrafted priors, offering moderate improvement. DIP (orange region) introduces implicit regularization and moves closer to the solution, while DIP with additional priors (pink region) further improves accuracy. The proposed $D^2IP$ (blue) benefits from UPWS and TPP, starting closer and converging faster. Red and blue dots denote potential reconstructions.}
\label{solution_space}
\end{figure}

\begin{figure}[!t]
\centerline{\includegraphics[scale=0.27]{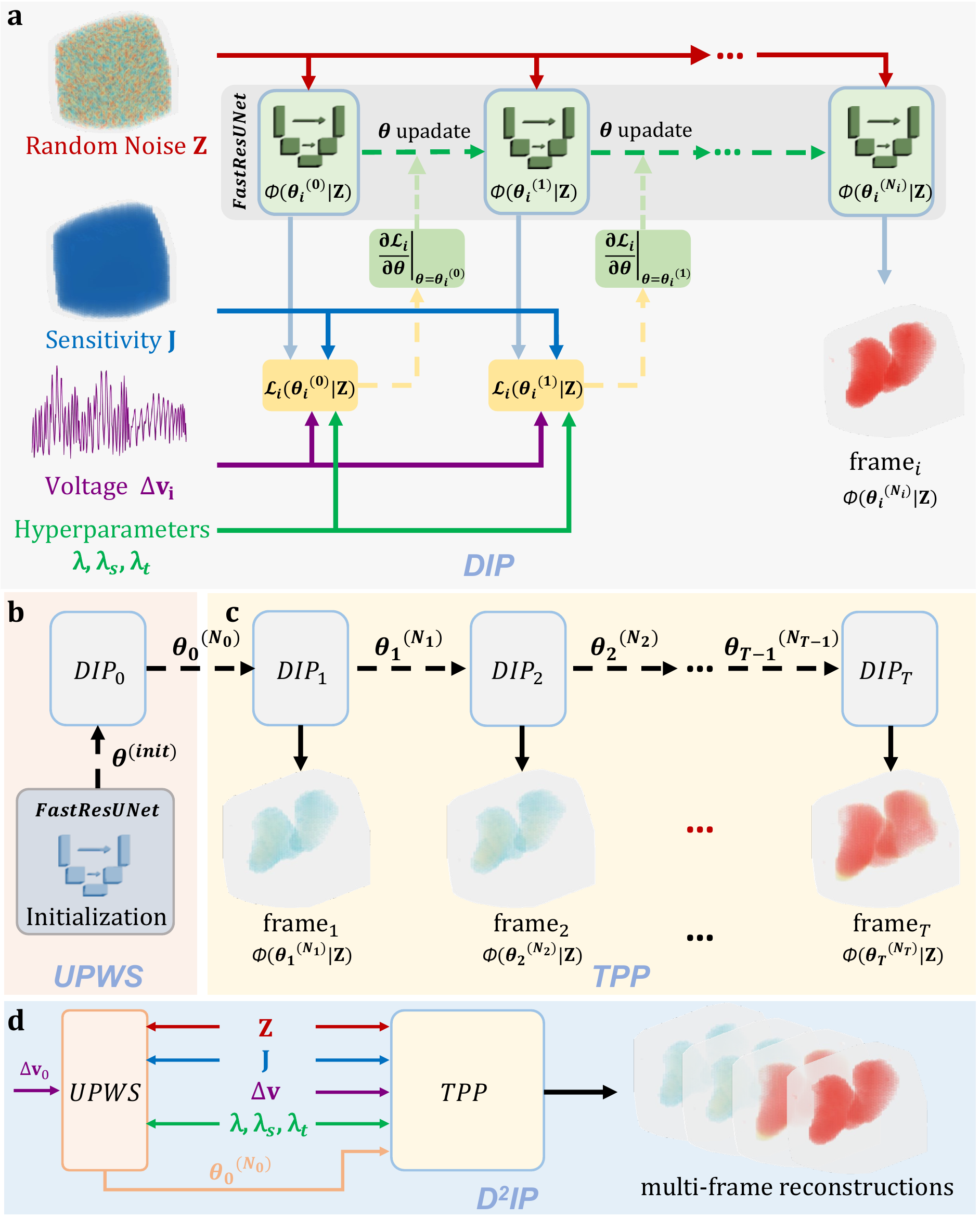}}
\caption{Illustration of the proposed $D^2IP$ framework. 
(a) DIP-based optimization using the \emph{3D-FastResUNet} with depthwise separable convolutions \cite{b45}. (b) UPWS first initializes the \emph{3D-FastResUNet} parameters using Kaiming initialization, followed by an iterative optimization process on an arbitrary single-frame reconstruction task. (c) TPP propagates learned parameters across sequential frames, ensuring temporal consistency in 3D time-sequence reconstruction. (d) Full $D^2IP$ pipeline integrating UPWS and TPP for multi-frame reconstruction.
}
\label{DDIP}
\end{figure}

As illustrated in Fig.~\ref{solution_space}, the proposed $D^2IP$ integrates both explicit and implicit priors via UPWS and TPP strategies, achieving superior initialization and convergence compared to traditional methods. Fig.~\ref{DDIP} provides an overview of the framework. In $D^2IP$, the unknown conductivity distribution is represented as the output of an untrained neural network, leveraging implicit regularization for reconstruction. Within our $D^2IP$ framework, the unknown conductivity distribution is expressed as the output of our \emph{3D-FastResUNet} $\phi: \mathbb{R}^{R\times C\times P} \to \mathbb{R}^{R\times C\times P}$, i.e.,
\begin{equation}
\Delta\pmb{\sigma}_i=\mathcal{V}\left(\phi\left(\pmb{\theta}_{i}^{(k)}|\mathbf{Z}\right)\right),
\label{eq:DDIP}
\end{equation}
where $\pmb{\theta}_{i}^{(k)}$ denotes the parameters of \emph{3D-FastResUNet} at the $k$-th iteration for reconstructing the $i$-th conductivity image, and $\mathbf{Z}\in\mathbb{R}^{R\times C\times P}$ is an input noise tensor sampled from a uniform distribution $\mathbf{Z}\sim U(0,1)$. The operator $\mathcal{V}: \mathbb{R}^{R\times C\times P} \to \mathbb{R}^{(R\times C \times P)\times 1}$ reshapes the output tensor from a 3D conductivity distribution to a 1D representation for optimization.

By substituting \eqref{eq:DDIP} into \eqref{eq:argmin}, the reconstructed conductivity distribution $\Hat{\Delta\pmb{\sigma}_i}$ can be further expressed as:
\begin{align}
\Hat{\Delta\pmb{\sigma}_i} = \arg\min_{\Delta\pmb{\sigma}_i} & \quad \|\Delta\mathbf{v}_i - \mathbf{J} \mathcal{V}\left(\phi\left(\pmb{\theta}_{i}|\mathbf{Z}\right)\right)\|  \notag \\
&+ \lambda\mathcal{R}\left(\{\left(\Hat{\Delta\pmb{\sigma}_j}\right)\}_{j=1}^{i-1},\mathcal{V}(\phi\left(\pmb{\theta}_{i}|\mathbf{Z}\right))\right).
\label{eq:DDIPargmin}
\end{align}




Here, the $\ell_2$ norm is used for data fidelity. The input for the regularization term $\mathcal{R}$ includes all previously reconstructed frames $\{\left(\Hat{\Delta\pmb{\sigma}_i}\right)\}_{j=1}^{i-1}$ and the current frame $\mathcal{V}(\phi\left(\pmb{\theta}_{i}|\mathbf{Z}\right))$. The prevailing Adam \cite{b42} optimizer 
is employed to optimize (\ref{eq:DDIPargmin}), ensuring stable and efficient parameter updates of \emph{3D-FastResUNet}. The optimized conductivity at the final iteration is given by:
\begin{equation}
\Hat{\Delta\pmb{\sigma}_i}=\mathcal{V}\left(\phi\left(\pmb{\theta}_{i}^{(N_{i})}|\mathbf{Z}\right)\right),
\end{equation}
where $N_{i}$ denotes the maximum number of iterations for reconstructing the $i$-th frame. Note that the iteration numbers for our tsEIT pulmonary imaging experiments among the UPWS phase, the first-frame reconstruction, and the subsequent consecutive reconstructions are set to 1800:450:250.

\begin{figure*}[!t]
\centerline{\includegraphics[scale=0.25]{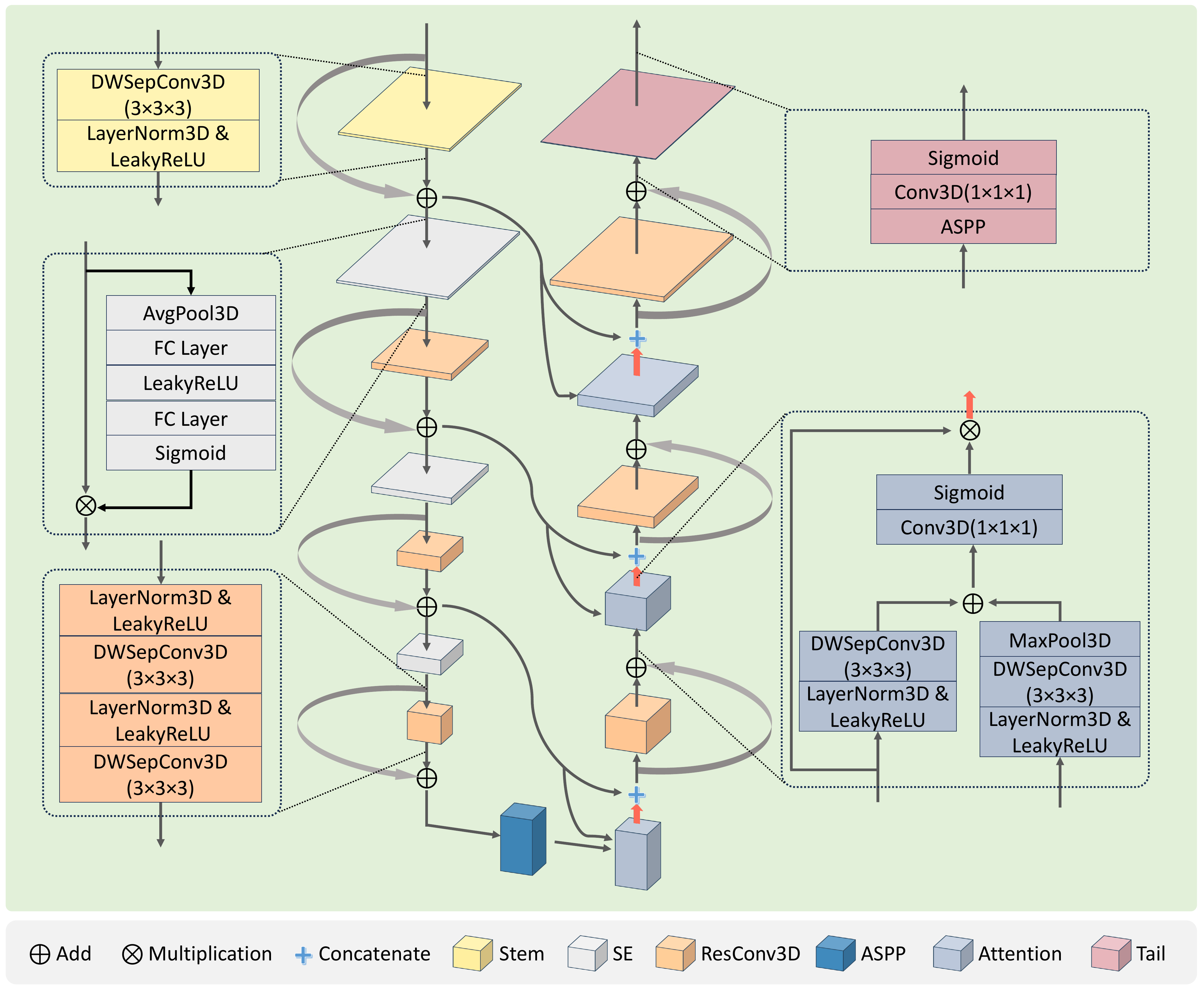}}
\caption{The architecture of the proposed \emph{3D-FastResUNet}.}
\label{network}
\end{figure*}

\subsection{Unsupervised Parameter Warm-Start}
To accelerate the convergence of $D^2IP$, we introduce an Unsupervised Parameter Warm-Start (UPWS) strategy, as illustrated in Fig.~\ref{DDIP}b. UPWS first initializes the \emph{3D-FastResUNet} parameters using Kaiming initialization \cite{b55}, followed by an iterative optimization process on an arbitrary single-frame reconstruction task. The optimized network parameters are then transferred as initialization for subsequent reconstructions, significantly reducing the required number of iterations (see Fig.~\ref{convergence_analysis}b and Fig.~\ref{inference_time}). In practical scenarios, the final parameters from the previous multi-frame reconstruction task can also be directly reused to warm-start the new reconstruction.

Formally, given a single-frame voltage measurement $\Delta \mathbf{v}_{0}$ (see Fig.~\ref{DDIP}b) — which can be selected from any measurement acquired under the same experimental protocol — the pretraining phase optimizes the network parameters $\pmb{\theta}$ by solving:
\begin{equation}
\pmb{\theta}_{0}^{(N_0)} = \arg\min_{\pmb{\theta}_{0}} \|\Delta\mathbf{v}_{0} - \mathbf{J} \mathcal{V}(\phi(\pmb{\theta}_{0} | \mathbf{Z}))\|+\lambda\mathcal{R}\left(\mathcal{V}\left(\phi(\pmb{\theta}_{0} | \mathbf{Z})\right)\right),
\label{eq:upws_single}
\end{equation}
where $N_0$ denotes the total number of iterations in the UPWS stage, determined empirically from the convergence curves shown in Fig.~\ref{convergence_analysis}a. The resulting parameters $\pmb{\theta}^{(N_0)}$ are then used to initialize the reconstruction of the first time-sequence frame:
\begin{equation}
\pmb{\theta}_{1}^{(0)} = \pmb{\theta}_{0}^{(N_0)}.
\label{eq:upws_transfer}
\end{equation}
Accordingly, the conductivity distribution for the first frame is obtained via:
\begin{equation}
\Hat{\Delta\pmb{\sigma}_{1}} = \mathcal{V}\left( \phi\left(\pmb{\theta}_{1}^{(N_{1})} \mid \mathbf{Z} \right) \right),
\label{eq:tpp_output2}
\end{equation}


\subsection{Temporal Parameter Propagation}
To further accelerate convergence and enforce temporal consistency across sequential frames, we introduce the Temporal Parameter Propagation (TPP) strategy. Instead of optimizing each frame independently, TPP transfers the optimized parameters of the previous frame as initialization for the next frame. This strategy leverages temporal correlations between adjacent frames, effectively reduces the number of iterations required for each subsequent reconstruction (see Fig.~\ref{convergence_analysis}b and Fig.~\ref{inference_time}), while preserving the temporal coherence of the reconstructed conductivity (see Fig.~\ref{realworld_comparison}).

Formally, given the optimized parameters $\pmb{\theta}_{i}^{(N_{i})}$ from the reconstruction of the $i$-th frame, the initial parameters for the $(i+1)$-th frame are set as:
\begin{equation}
\pmb{\theta}_{i+1}^{(0)} = \pmb{\theta}_{i}^{(N_{i})},
\label{eq:tpp_transfer}
\end{equation}
The final conductivity distribution for the $(i+1)$-th frame is then generated by passing the fixed input noise $\mathbf{Z}$ through the optimized network:
\begin{equation}
\Hat{\Delta\pmb{\sigma}_{i+1}} = \mathcal{V}\left( \phi\left(\pmb{\theta}_{i+1}^{(N_{i+1})} \mid \mathbf{Z} \right) \right),
\label{eq:tpp_output}
\end{equation}


\subsection{3D-FastResUNet}
The \emph{3D-FastResUNet} $\phi: \mathbb{R}^{R\times C\times P} \to \mathbb{R}^{R\times C\times P}$ is specifically designed for $D^2IP$ to serve as an implicit image prior. Unlike conventional DIP methods that typically employ 2D UNet structures, we adopt 3D convolutions throughout the network to better preserve spatial continuity and volumetric consistency in 3D reconstruction tasks. The network architecture is illustrated in Fig.~\ref{network}. It builds upon \cite{b15} by extending the architecture to 3D convolutional operations and introducing depthwise separable convolutions (DWSepConv3D) \cite{b45}, which significantly enhance spatial representation capability and computational efficiency. Additionally, residual connections and LayerNorm are retained to ensure stable convergence without requiring specially designed stopping criteria.

The \emph{3D-FastResUNet} consists of several key components, i.e., a stem block, six 3D residual convolutional (ResConv3D) blocks, three Squeeze-and-Excitation (SE) \cite{b43} units, an Atrous Spatial Pyramid Pooling (ASPP) \cite{b44} module, three attention units, and a tail block. The network begins with a stem block, followed sequentially by three pairs of SE units and ResConv3D blocks. At the bottleneck, we incorporate an ASPP module to increase the representational capacity of the network by introducing multi-scale receptive fields, which strengthens the ability of the network to model complex spatial structures from the noise input. The decoder path progressively upsamples the latent features and fuses them with skip connections from the encoder. Each upsampling stage includes an attention mechanism, followed by interpolation and a ResConv3D block, ensuring spatial coherence across scales. Finally, the output is passed through a tail block comprising a Conv3D layer and a sigmoid activation to generate the reconstructed 3D conductivity distribution. 

\begin{figure}[!t]
\centerline{\includegraphics[scale=0.3]{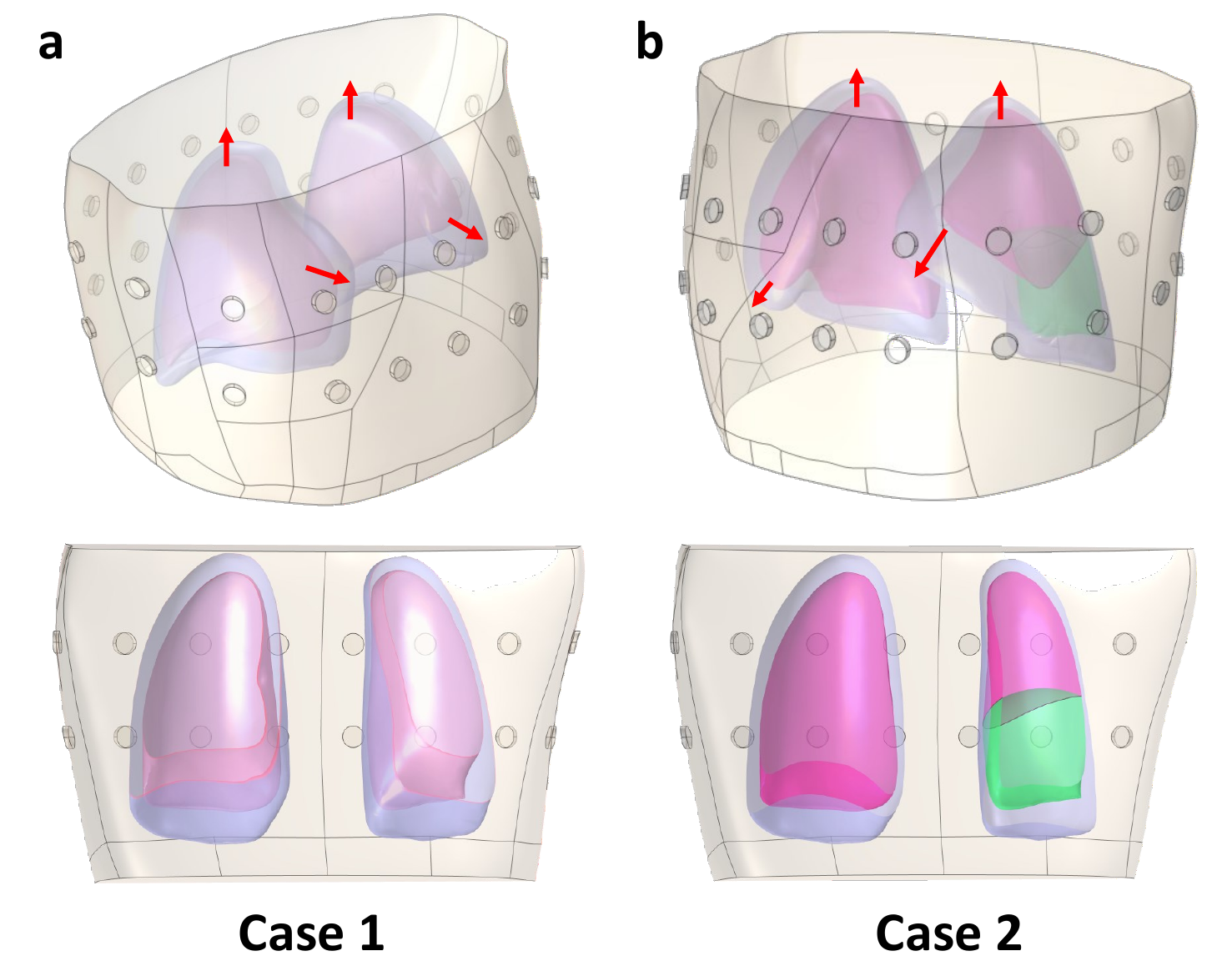}}
\caption{Illustration of the two simulated lung expansion scenarios: (a) Lung expansion over a deep breathing cycle in a healthy thoracic model. (b) Lung expansion in the presence of pulmonary edema in the left lung during a deep breathing cycle.}
\label{Simulation_exp}
\end{figure}

\subsection{4D Total Variation}
Given the ill-posed nature of EIT, DIP-based reconstructions without regularization often exhibit instability and noise amplification \cite{b35,b36}, as evidenced by the divergence trends shown in Fig.~\ref{convergence_analysis}b. Therefore, we introduce a 4D Total Variation (4D-TV) regularization term as a critical component to enforce spatial-temporal coherence and improve reconstruction robustness. In our formulation, spatial TV is computed on the current frame $\Delta\pmb{\sigma}_i$, while temporal TV is applied across all previous frames $\{\Hat{\Delta\pmb{\sigma}_j}\}_{j=1}^{i-1}$ and the current frame to ensure temporal coherence throughout the sequence. The separation of spatial and temporal components in 4D-TV allows for tailored regularization strengths, enabling adaptive regularization, thereby improving reconstruction stability and preserving dynamic structural changes. The 4D-TV regularization is defined as:
\begin{equation}
\begin{aligned}
&\mathcal{R}_{\text{4D-TV}}(\{\Hat{\Delta\pmb{\sigma}_j}\}_{j=1}^{i-1},\Delta\pmb{\sigma}_i) \\
&=
\lambda_s \, \mathcal{R}_{\text{spatial}}(\Delta\pmb{\sigma}_i) +\lambda_t \, \mathcal{R}_{\text{temporal}}(\{\Hat{\Delta\pmb{\sigma}_j}\}_{j=1}^{i-1},\Delta\pmb{\sigma}_i),
\end{aligned}
\end{equation}
where $\lambda_s$ and $\lambda_t$ are weighting coefficients for the spatial and temporal terms, respectively. The spatial and temporal regularizations are given by:
\begin{equation}
\begin{aligned}
\mathcal{R}_{\text{spatial}}(\Delta\pmb{\sigma}_i) = \frac{1}{V}
\sum_{p,r,c} 
\sqrt{
\begin{aligned}
& \|\Delta\pmb{\sigma}_{i,p+1,r,c} - \Delta\pmb{\sigma}_{i,p,r,c} \|^2 + \\
& \|\Delta\pmb{\sigma}_{i,p,r+1,c} - \Delta\pmb{\sigma}_{i,p,r,c} \|^2 + \\
& \|\Delta\pmb{\sigma}_{i,p,r,c+1} - \Delta\pmb{\sigma}_{i,p,r,c} \|^2 + \epsilon
\end{aligned}
}.
\end{aligned}
\end{equation}

\begin{equation}
\begin{aligned}
&\mathcal{R}_{\text{temporal}}(\{\Hat{\Delta\pmb{\sigma}_j}\}_{j=1}^{i-1},\Delta\pmb{\sigma}_i) \\
&= 
\frac{1}{\sum_{j=1}^{i-1} \alpha_{j,i}}\sum_{j=1}^{i-1} \alpha_{j,i} \sum_{p,r,c} \sqrt{ 
(\Delta\pmb{\sigma}_{i,p,r,c} - \Hat{\Delta\pmb{\sigma}_{j,p,r,c}})^2 + \epsilon }.
\end{aligned}
\end{equation}
where $p$, $r$, and $c$ index the depth, height, and width of each 3D conductivity distribution. $\epsilon$ is a small constant introduced to prevent division by zero. $\alpha_{j,i} = \exp(-(i - j))$ is a temporal decay factor that controls the influence of previous frames, giving more weight to recent frames while allowing flexibility for long-term structural evolution.

The 4D-TV regularization is incorporated into the overall loss function as:
\begin{equation}
\begin{aligned}
\mathcal{L}_{i} = 
& \quad \|\Delta\mathbf{v}_i - \mathbf{J} \mathcal{V}(\phi(\pmb{\theta}_i | \mathbf{Z}))\| \\
& + \lambda_{\text{TV}} \mathcal{R}_{\text{4D-TV}}(\{\Hat{\Delta\pmb{\sigma}_j}\}_{j=1}^{i-1},\mathcal{V}(\phi(\pmb{\theta}_i | \mathbf{Z}))),
\end{aligned}
\label{eq:loss_4dtv}
\end{equation}
where $\lambda_{\text{TV}}$ controls the weight of the total regularization term.

\section{Experimental Setup}
\subsection{Simulation Setup and Synthetic Data Generation}
We employ COMSOL Multiphysics to simulate two groups of 3D time-sequence Electrical Impedance Tomography (tsEIT) measurements in a thoracic model, as illustrated in Fig.~\ref{Simulation_exp}. The simulation domain incorporates a 3D lung model extracted from real CT scans, providing anatomically realistic representations of both the thoracic boundary and internal lung structures and anatomical changes in lung geometry between the beginning and end of a deep inhalation. To simulate intermediate lung expansion states throughout the breathing cycle, we use interpolation between these two anatomically accurate endpoints. The interpolation is performed by simple isotropic scaling relative to the geometric center of the lung, with the lung volume varying sinusoidally to align with real physiological dynamics.
In both experiments, the background conductivity is set to 0.24~S/m, and 32 electrodes are evenly distributed in two layers along the thoracic boundary.

\begin{figure*}[!t]
\centerline{\includegraphics[scale=0.51]{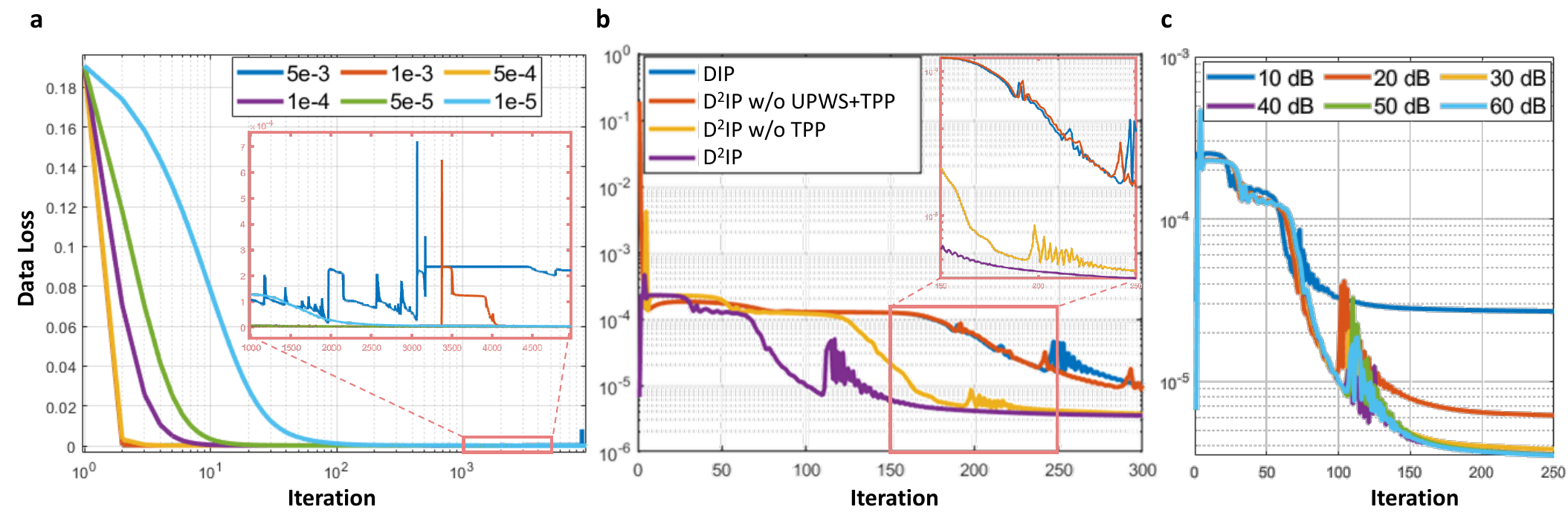}}
\caption{Convergence analysis of $D^2IP$  under different settings. (a) Effect of learning rate on $D^2IP$  convergence. (b) Ablation study of acceleration strategies including UPWS and TPP. (c) Convergence performance of $D^2IP$  under different levels of input noise (10–60 dB).}
\label{convergence_analysis}
\end{figure*}

\begin{figure*}[!t]
\centerline{\includegraphics[scale=0.27]{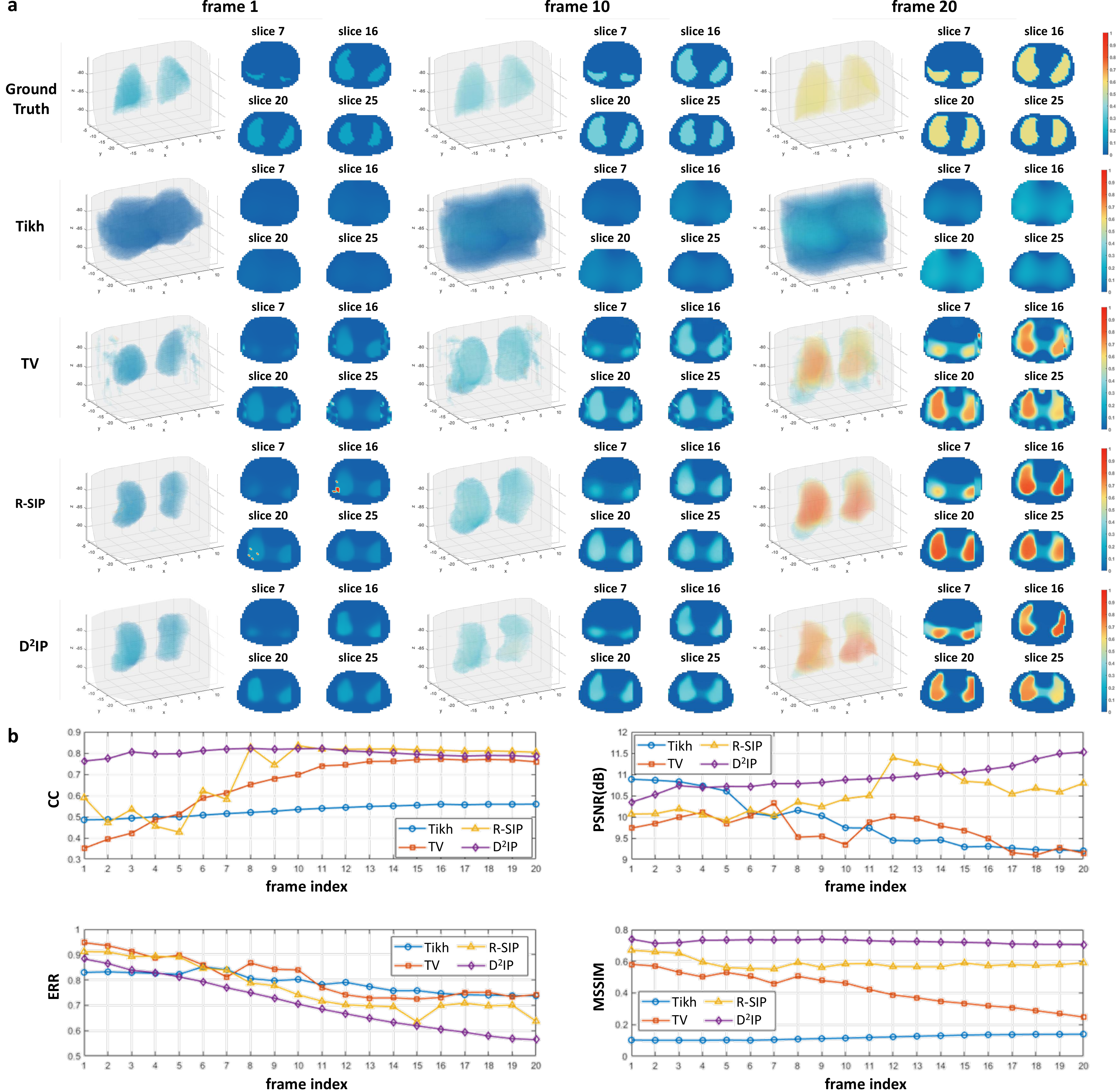}}
\caption{Qualitative and quantitative comparison of different reconstruction methods on Case 1. (a) Visual comparison of reconstructed frames (frames 1, 10, and 20) using different reconstruction methods, alongside the ground truth. (b) Quantitative evaluation across the full time sequence in terms of CC, PSNR, ERR, and MSSIM metrics.}
\label{Simulation_comparison_exp1}
\end{figure*}

\begin{figure}[!t]
\centerline{\includegraphics[scale=0.3]{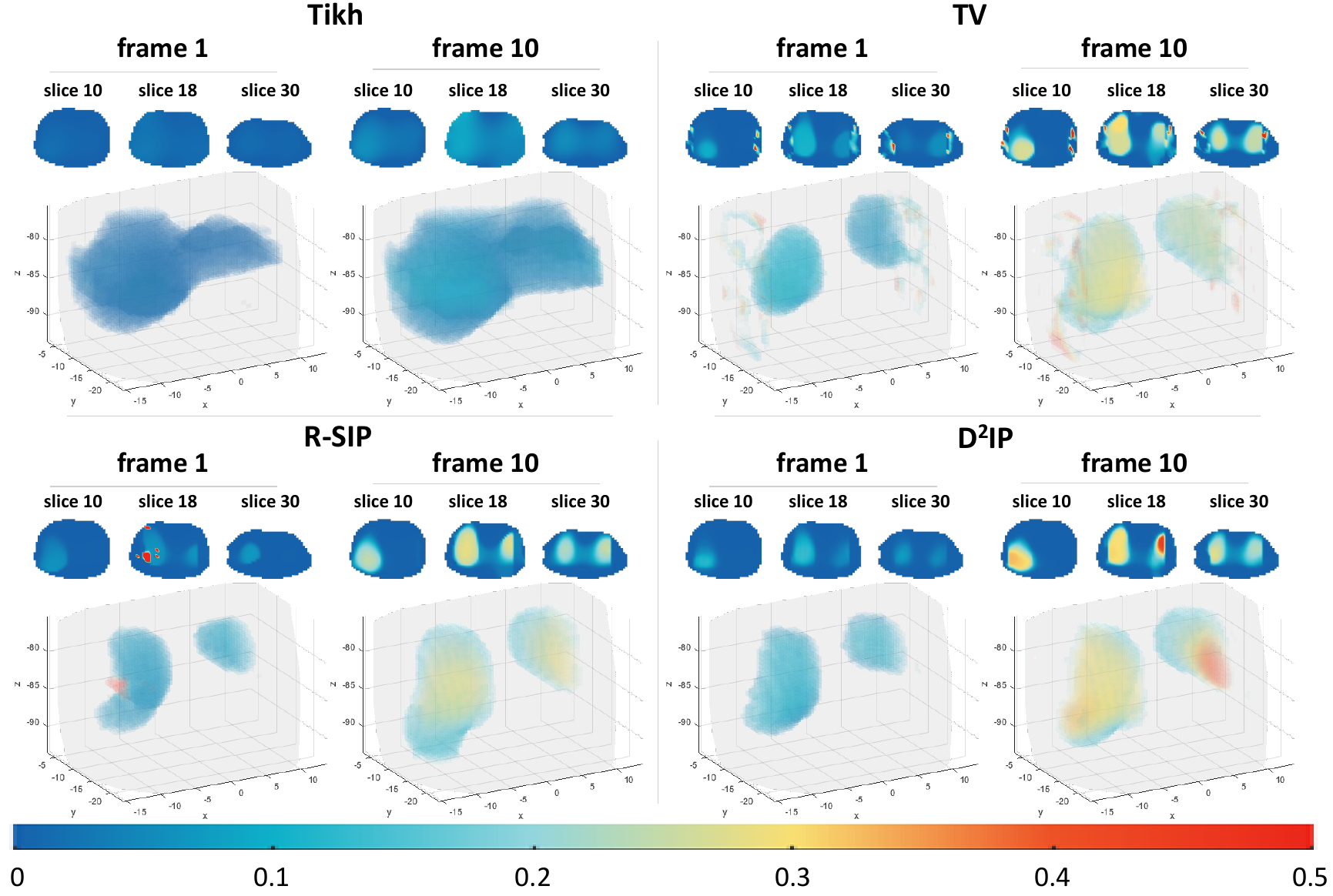}}
\caption{Visual comparison of reconstructed frames (frames 1, 10) using different reconstruction methods on Case 2.}
\label{Simulation_comparison_exp2}
\end{figure}

\begin{figure}[!t]
\centerline{\includegraphics[scale=0.42]{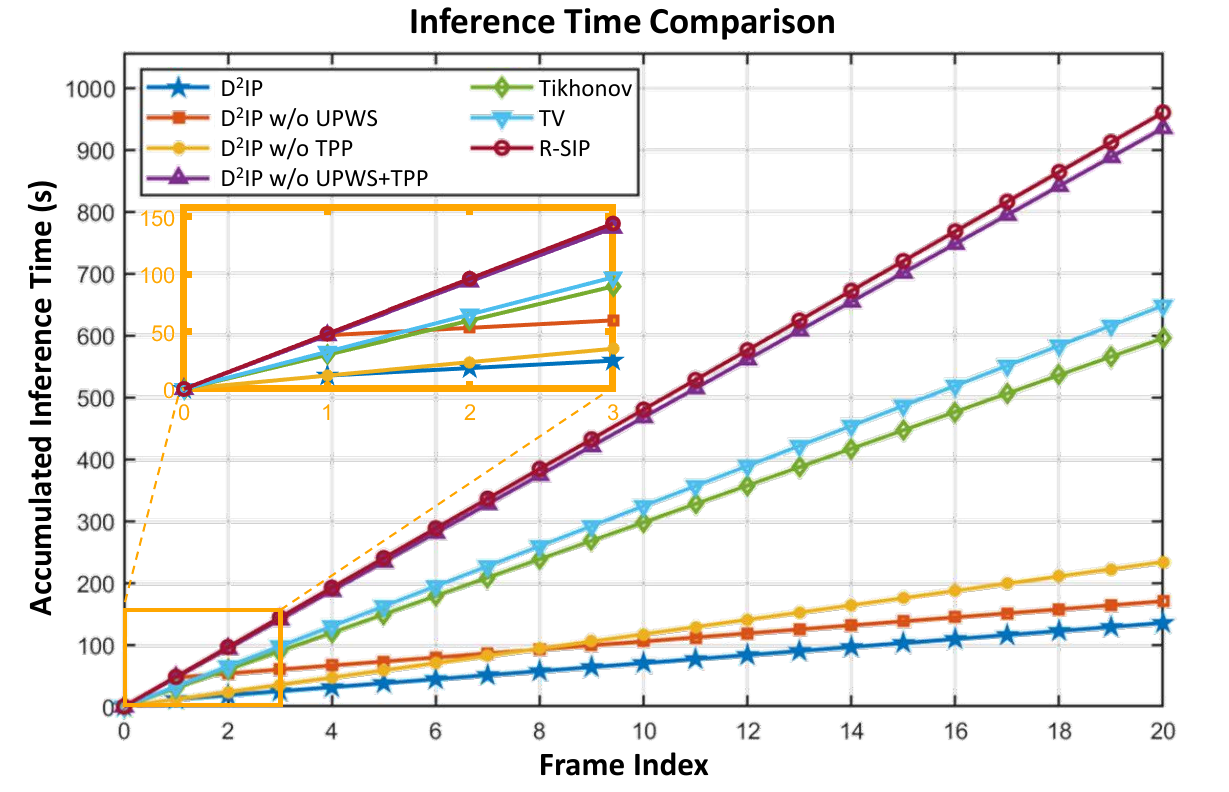}}
\caption{Comparison of accumulated inference time across 20 frames for different reconstruction methods.}
\label{inference_time}
\end{figure}

\begin{figure*}[!t]
\centerline{\includegraphics[scale=0.35]{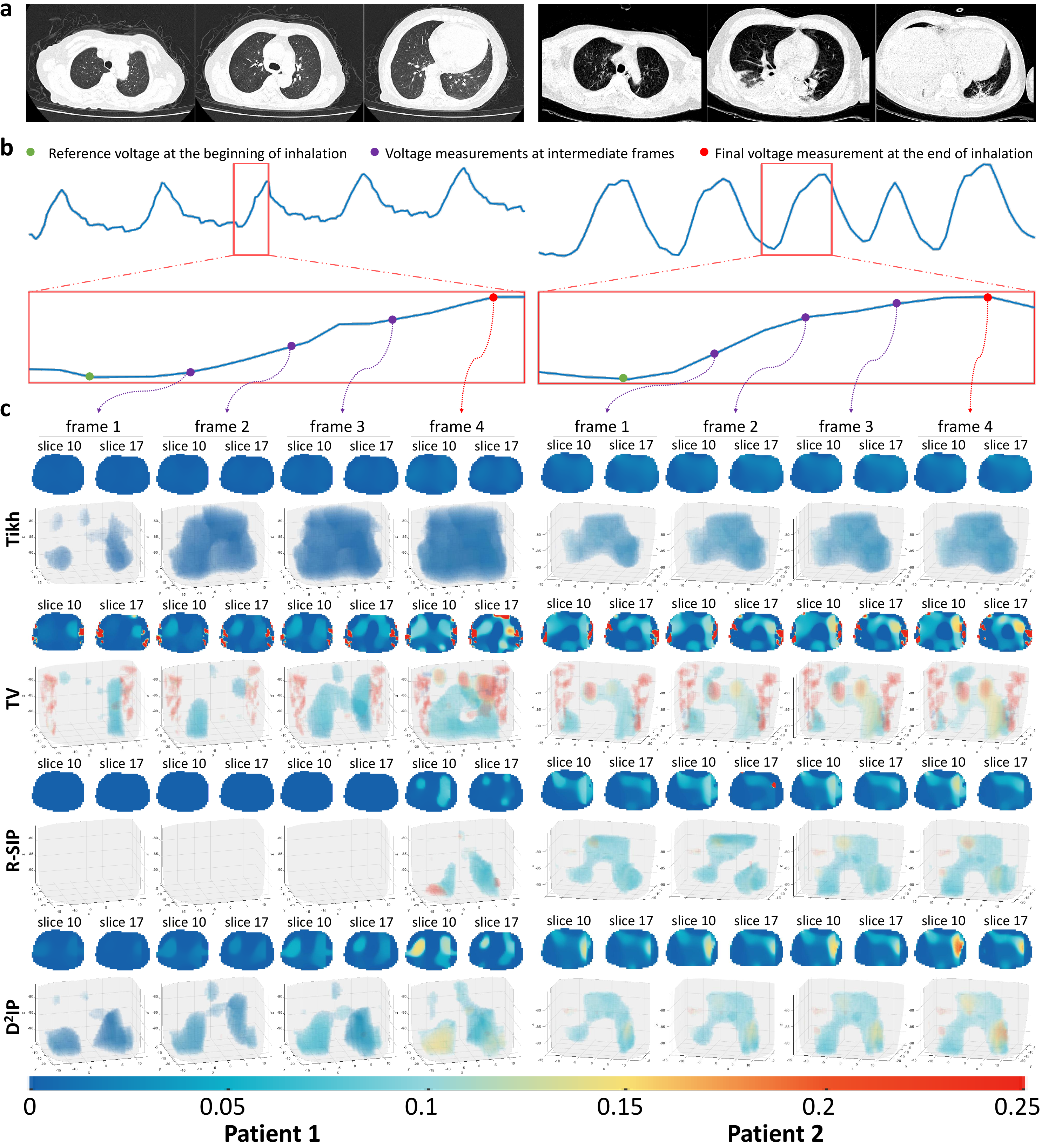}}
\caption{Real-world clinical experiment setup and reconstruction results. (a) Axial CT images of two participants (viewed from the feet upwards). (b) Top: the sum of absolute EIT voltage measurements over five breathing cycles, reflecting global impedance changes over time. Bottom: a selected full inhalation phase used for 3D time-sequence EIT reconstruction, with marked frame positions. (c) tsEIT reconstruction results at multiple frames during the selected inhalation phase, with axial slices (10, 17) and 3D views for both patients.}
\label{realworld_comparison}
\end{figure*}

The first simulation case (Fig.~\ref{Simulation_exp}(a)) models normal lung expansion during a deep breathing cycle in a healthy subject. The conductivity of the lung tissue gradually decreases from 0.20~S/m at the start of inhalation to 0.105~S/m at full expansion, simulating increased air content. The empty thoracic background is used as the reference frame for voltage calculation. The second case (Fig.~\ref{Simulation_exp}(b)) models a tidal breathing process in the presence of pulmonary edema in the left lung. In this case, the healthy lung regions undergo normal conductivity changes, decreasing from 0.14~S/m to 0.0835~S/m throughout the breathing cycle. The edematous region of the left lung maintains a constant conductivity of 0.24~S/m, equivalent to the background. In this case, the first frame is used as the reference to compute the differential voltages.

Each breathing sequence is simulated over 20 frames, and for each frame, 328 boundary voltage measurements are generated using the measurement protocol described in~\cite{b36}. The first experiment yields a time sequence of $328 \times 20$ normalized voltage measurements, while the second produces $328 \times 19$ normalized voltage measurements as the first frame is used as a reference.

\subsection{Clinical Data Acquisition and Experimental Protocol}
To evaluate the effectiveness of $D^2IP$  under practical clinical conditions, we conducted an experiment involving two participants (denoted as Patient 1 and Patient 2). The study protocol was approved by the Institutional Review Board (IRB) of Peking Union Medical College Hospital (Approval No. I-24PJ1732). Written informed consent was obtained from both participants prior to any data collection, using the IRB-approved consent forms (Version 2.0, dated July 22, 2024). All procedures adhered to the Declaration of Helsinki and national ethical regulations, with participant data anonymized to ensure confidentiality.

Before each EIT measurement session, chest CT scans were acquired (see Fig.~\ref{realworld_comparison}a) to serve as anatomical references for subsequent medical image analysis. A dual-layer thoracic electrode belt with 32 electrodes uniformly distributed across two horizontal planes was used to acquire 3D time-sequence EIT voltage measurements. The participants remained seated during the entire procedure and followed a controlled breathing protocol. Voltage data were collected using the same measurement protocol as the simulation. Patient 1 exhibited normal bilateral lung ventilation, while Patient 2 presented with right lung infection and corresponding ventilation impairment.

\subsection{Comparison Algorithms}
We compare $D^2IP$  against three baselines: Tikhonov (Tikh) regularization \cite{b9}, Total Variation (TV) regularization \cite{b11}, and Regularized Shallow Image Prior (R-SIP) \cite{b36}. Tikh is a classical model-based single-step method for EIT reconstruction. Its 3D version can involve significant computational costs due to large-scale matrix operations. TV corresponds to the spatial component $\mathcal{R}_{\text{spatial}}(\cdot)$ of the 4D Total Variation (4D-TV) used in $D^2IP$ . R-SIP represents the state-of-the-art (SOTA) unsupervised approach for 3D EIT image reconstruction, combining handcrafted regularization with a shallow neural network to enhance robustness and performance.

\subsection{Parameter Settings}
The parameter settings for $D^2IP$ are determined based on convergence curve analysis, quantitative evaluation metrics, and empirical tuning through trial-and-error. The number of iterations is set to 1,800 for the warm-start phase, 450 for the first-frame reconstruction, and 250 for each subsequent frame. The learning rate is set to $5\times10^{-4}$ for all simulation experiments and $1\times10^{-4}$ for real-world experiments. For the 4D-TV regularization, $\lambda$ is set to 0.002, with spatial and temporal weighting parameters $\lambda_s=1$ and $\lambda_t=0.1$, respectively, for all of the experiments. 

For comparison methods, we adopt the optimal parameter settings reported in \cite{b36} to ensure fair comparisons for TV and R-SIP. For Tikh regularization, the regularization parameter $\mu$ is selected by searching within the range of 0.001 to 0.01, and the value yielding the best reconstruction performance is adopted. All experiments are conducted under identical reconstruction settings, ensuring a fair and consistent evaluation.

\section{Results and discussion}

\subsection{Simulation Data}
\subsubsection{Convergence Behavior Analysis}
We evaluate the convergence behavior of $D^2IP$ using Case 1. As shown in Fig.~\ref{convergence_analysis}, we investigate the effects of the learning rate, acceleration strategies, and noise level on convergence performance.

Fig.~\ref{convergence_analysis}(a) illustrates the data loss variation over 10{,}000 iterations under different learning rates. This experiment is conducted using the first voltage frame of Case 1 with randomly initialized network parameters, without applying the UPWS or TPP acceleration strategies. The results show that excessively large learning rates (e.g., 5e-3 and 1e-3) lead to significant oscillations, while smaller learning rates (e.g., 1e-4, 5e-5, and 1e-5) yield smoother convergence at the cost of slower speed. Considering the trade-off between stability and efficiency, we set the learning rate to $5\times10^{-4}$ to ensure both fast and stable convergence. Under this setting, the optimization typically reaches a plateau around 1,500 iterations; therefore, we choose 1,800 iterations for the warm-start phase to provide sufficient optimization margin while avoiding overfitting.

Fig.~\ref{convergence_analysis}(b) presents an ablation study on the impact of acceleration strategies, based on the second voltage frame of Case 1. The blue curve corresponds to the baseline DIP, which excludes acceleration strategies and handcrafted priors. The red curve incorporates handcrafted regularization. The yellow curve represents the UPWS strategy, where the network is first optimized on a random reconstruction task for 1,800 iterations before being transferred to the target frame. The purple curve denotes the full $D^2IP$, where the network is initialized using the parameters from the first-frame reconstruction. The results confirm that both UPWS and TPP significantly accelerate convergence: the full $D^2IP$  achieves stable convergence within approximately 200 iterations, while $D^2IP$  without TPP also converges within around 300 iterations. Based on these observations, we set the number of iterations to 450 for the first-frame reconstruction, and 250 for each subsequent frame. These values include an additional margin beyond the initial convergence points to ensure stable optimization while mitigating the risk of overfitting.

Fig.~\ref{convergence_analysis}(c) evaluates the robustness of $D^2IP$  under noisy conditions. Gaussian noise is added to the voltage measurements to simulate different signal-to-noise ratios (SNRs) ranging from 10~dB to 60~dB. Across all noise levels, $D^2IP$  maintains rapid and stable convergence, typically within 200 iterations. Interestingly, we observed that the data loss curve exhibits noticeable fluctuations during early iterations in noise-free simulation data, with similar but slightly reduced behavior under high-SNR conditions. In contrast, low-SNR and real measured data consistently yield smoother loss curves from the beginning. This phenomenon arises primarily due to the high expressiveness of the \emph{3D-FastResUNet} architecture: in noise-free or very high-SNR cases, the network tends to aggressively fit subtle structural details, resulting in sensitivity and fluctuations amplified by the element-wise sensitivity of the measurement-domain MSE loss. Conversely, higher noise levels encourage the network to initially focus on robust, low-frequency features, naturally smoothing out early-stage optimization gradients. Nonetheless, regardless of noise conditions, as training progresses and the network learns dominant image features, the optimization curves universally converge toward a stable and smooth pattern.

\subsubsection{Qualitative and Quantitative Reconstruction Results}

Fig.~\ref{Simulation_comparison_exp1}a shows the qualitative reconstruction results of different algorithms for Case 1. We visualize three representative frames across the breathing cycle: frame 1 (beginning of inhalation), frame 10 (middle of inhalation), and frame 20 (end of inhalation). For each frame, both 3D renderings and axial slices at multiple depths (slices 7, 16, 20, and 25) are shown to provide a comprehensive assessment of spatial fidelity.

In the early phase of inhalation (frame 1), where conductivity changes are minimal, the Tikhonov method fails to produce meaningful structures. TV and R-SIP successfully recover the general lung shape but exhibit noticeable noise artifacts. In contrast, the proposed $D^2IP$  method reconstructs clean and well-defined lung contours without apparent noise. At the middle frame (frame 10), as the conductivity variation increases, Tikhonov begins to reveal vague outlines of the lung structure. TV produces clearer contours but still suffers from substantial noise, especially in deeper slices. Both R-SIP and $D^2IP$  deliver clean reconstructions; however, $D^2IP$  demonstrates superior structure preservation. By the end of inhalation (frame 20), where the conductivity contrast between lung and background is most pronounced, all methods are able to differentiate the left and right lungs. While TV, R-SIP, and $D^2IP$  generate visually clean reconstructions, $D^2IP$  again shows better edge sharpness and spatial detail.

Fig.~\ref{Simulation_comparison_exp1}b presents the quantitative evaluation across 20 frames using four metrics: correlation coefficient (CC), peak signal-to-noise ratio (PSNR), mean structural similarity (MSSIM), and error ratio (ERR). These metrics jointly assess reconstruction accuracy, structural fidelity, and robustness to noise. The proposed $D^2IP$  achieves the highest CC values during the first ten frames, and remains highly competitive in later frames, slightly trailing R-SIP after frame 10. For ERR, $D^2IP$  maintains the lowest values across most of the sequence, with the exception of the first two frames where Tikh achieves marginally lower errors. In terms of MSSIM, $D^2IP$  consistently outperforms all baselines, reflecting its strong capability in preserving structural similarity over time. PSNR results show that $D^2IP$  achieves the most stable performance, with the highest values in the later frames (14–20). It is slightly lower than Tikh in the first three frames and marginally outperformed by R-SIP around frames 11–13. In addition to frame-wise performance, we further compared the average values of different metrics across all 20 frames. $D^2IP$ achieves the highest mean CC (0.80), PSNR (10.92), and MSSIM (0.7332), while also maintaining the lowest average ERR (0.7067).

Fig.~\ref{Simulation_comparison_exp2} shows the reconstruction results of different algorithms on Case 2. Since this case is based on time difference imaging, no ground truth conductivity distribution is available for quantitative evaluation. Therefore, we focus on qualitative visual assessment to compare the structural fidelity and noise robustness of different methods. Two representative frames are selected for visualization: frame 1 (beginning of inhalation) and frame 10 (mid-inhalation). At frame 1, both TV and R-SIP exhibit noticeable noise artifacts under low conductivity contrast conditions, consistent with observations in Case 1. At frame 10, all methods successfully reconstruct the ventilation-induced conductivity changes in the healthy right lung, as seen in slice 10, while showing minimal response in the edematous left lung due to its lack of conductivity variation. 


Both qualitative and quantitative results across Case 1 and Case 2 demonstrate the superior performance of $D^2IP$ in reconstructing dynamic conductivity distributions. $D^2IP$  consistently yields cleaner images with sharper anatomical boundaries and reduced noise, particularly under low-contrast conditions or at physiological transition interfaces. Quantitatively, it achieves leading performance across key metrics including CC, ERR, MSSIM, and PSNR, highlighting its robustness and accuracy.

\subsubsection{Inference Time Comparison}
We compare the computational efficiency of all methods by reporting the accumulated inference time over a 20-frame 3D tsEIT reconstruction task, as shown in Fig.~\ref{inference_time}. All experiments were performed under identical hardware conditions (NVIDIA GeForce RTX 4090 GPU and Intel Core i9-14900KF CPU). Among all methods, $D^2IP$ achieves the fastest inference time (135.02s) among all compared methods over 20 frames, reducing runtime by 79.17\% compared to TV (648.21s), 85.94\% compared to R-SIP (960.13s), and 77.32\% compared to Tikhonov (595.20s), thereby significantly outperforming the SOTA baselines.

\subsubsection{Ablation Study}
To further evaluate the contributions of the proposed acceleration strategies, we conducted an ablation study by selectively removing UPWS and TPP from $D^2IP$ . While convergence behavior has already been analyzed in Fig.~\ref{convergence_analysis}b, we additionally compare the inference time across different $D^2IP$ variants in Fig.~\ref{inference_time}. 

The results show that $D^2IP$  variants without UPWS and/or TPP exhibit noticeably increased runtime. In particular, $D^2IP$  w/o UPWS + TPP requires nearly nine times the inference time of the full $D^2IP$ , approaching that of R-SIP. This confirms that the proposed acceleration strategies not only enhance convergence but also contribute significantly to practical efficiency.

\subsection{Clinical Data}
Fig.~\ref{realworld_comparison}c shows reconstruction results for two participants across an inhalation cycle. Four consecutive reconstructed frames are visualized using axial slices (10 and 17) and 3D views to reveal dynamic conductivity changes during breathing.

For Patient~1, $D^2IP$  successfully reconstructs symmetric and gradually increasing conductivity in both lungs, consistent with normal ventilation. Tikhonov and TV also capture this trend; however, Tikhonov suffers from excessive oversmoothing, resulting in the loss of fine structural details, while TV reconstructions exhibit significant noise artifacts. Due to the limited representational capacity of its shallow network, R-SIP fails to reconstruct meaningful results in the first three frames. Although it produces a reasonable output in the fourth frame, the result still exhibits noticeable noise, and the failure in earlier frames leads to a loss of temporal coherence across the sequence. $D^2IP$  reconstructions for Patient~2 clearly reveal asymmetry between the left and right lungs: the left lung demonstrates normal ventilation patterns, while the right lung exhibits minimal conductivity change, consistent with impaired ventilation caused by infection. This disparity becomes more apparent in deeper slices (e.g., slice 17) as inhalation progresses. Tikhonov fails to reflect this pathological feature, and while TV and R-SIP provide a partial indication of the asymmetry, both suffer from poor temporal coherence and reduced robustness, with reconstructions affected by noticeable noise artifacts.

These results demonstrate that $D^2IP$ accurately reconstructs both global ventilation trends and localized ventilation impairments, consistent with CT findings. 
Compared to baseline methods, $D^2IP$  offers superior robustness, noise resilience, and temporal consistency, which reinforces its clinical applicability.

\section{Conclusion}
We presented Deep Dynamic Image Prior ($D^2IP$), the first unsupervised learning framework for 3D time-sequence tomographic imaging, and demonstrated its effectiveness in the context of time-sequence Electrical Impedance Tomography (tsEIT). By integrating three key strategies—Unsupervised Parameter Warm-Start (UPWS), Temporal Parameter Propagation (TPP), and \emph{3D-FastResUNet}, $D^2IP$  accelerates convergence, enhances temporal coherence, and achieves high-quality reconstructions without requiring training data. Experiments on both simulated and clinical datasets show that $D^2IP$  outperforms classical and learning-based baselines in spatial accuracy, temporal stability, and computational efficiency while effectively revealing physiologically meaningful ventilation dynamics. Future work will explore extending $D^2IP$  to other dynamic tomographic modalities, such as CT and PET, and expanding its application to cardiac and tissue imaging.

\bibliographystyle{IEEEtran}
\bibliography{reference}

\end{document}